\documentclass[12pt]{article}

\usepackage{times}

\DeclareFontFamily{OT1}{times}{}
\DeclareFontShape{OT1}{times}{m}{n }{ <-> ptmr }{}
\DeclareFontShape {OT1}{times}{bx}{n }{ <-> ptmb }{}
\DeclareFontShape {OT1}{times}{m }{it}{ <-> ptmri}{}
\DeclareFontShape {OT1}{times}{bx}{it}{ <-> ptmbi}{}
\usepackage{amsmath}
\usepackage{amsfonts}
\usepackage{amssymb}
\usepackage{latexsym}
\usepackage{graphics}

\begin{document}

\newtheorem{theo}{Theorem}[section]
\newtheorem{defi}[theo]{Definition}
\newtheorem{prop}[theo]{Proposition}
\newtheorem{corr}[theo]{Corollary}
\newtheorem{lemm}[theo]{Lemma}
\newtheorem{exam}[theo]{Example}

\newcommand{\CD}[2]{\ensuremath{\raisebox{-.6ex}{\scriptsize{$0$}}\:\!
        \mathbf{C}_{t}^{#1}\left[#2\right]}}
\newcommand{\FFD}[4]{\ensuremath{\raisebox{-.6ex}{\scriptsize{$#1$}}\!
        \raisebox{1ex}{\scriptsize{$t$}}\:\!\mathbf{F}_{#2}^{#3}[#4]}}
\newcommand{\myH}[1]{\ensuremath{H^{#1}([0,T])}}
\newcommand{\K}[3]{\ensuremath{K_{#1}^{(#2)}(#3)}}
\newcommand{\Kstar}[3]{\ensuremath{K_{#1}^{(#2)} \star q^{#3}}}
\newcommand{\myL}{\ensuremath{L^{2}([0,T])}}
\newcommand{\LFD}[3]{\ensuremath{\raisebox{-.6ex}{\scriptsize{$#1$}}\:\!
        \mathbf{D}_{t}^{#2}\left[#3\right]}}
\newcommand{\LFDm}[3]{\ensuremath{\raisebox{-.6ex}{\scriptsize{$#1$}}\:\!
        \mathbf{D}_{t^{-}}^{#2}\left[#3\right]}}
\newcommand{\LFI}[3]{\ensuremath{\raisebox{-.6ex}{\scriptsize{$#1$}}\:\!
        \mathbf{I}_{t}^{#2}[#3]}}
\newcommand{\mysum}{\ensuremath{\sum_{n=1}^{\infty}}}
\newcommand{\PHIm}[2]{\ensuremath{\Phi_{#1}^{-}(#2)}}
\newcommand{\PHIp}[2]{\ensuremath{\Phi_{#1}^{+}(#2)}}
\newcommand{\PSI}[2]{\ensuremath{\Psi_{#1}(#2)}}
\newcommand{\lfeta}[2]{\ensuremath{\raisebox{-.6ex}{\scriptsize{$#1$}}\:\!\eta_{t}^{#2}}}
\newcommand{\rfeta}[2]{\ensuremath{\raisebox{-.6ex}{\scriptsize{$t$}}\:\!\eta_{#1}^{#2}}}
\newcommand{\lfq}[2]{\ensuremath{\raisebox{-.6ex}{\scriptsize{$#1$}}\:\!q_{t}^{#2}}}
\newcommand{\RFD}[3]{\ensuremath{\raisebox{-.6ex}{\scriptsize{$t$}}\:\!
        \mathbf{D}_{#1}^{#2}\left[#3\right]}}
\newcommand{\RFDp}[3]{\ensuremath{\raisebox{-.6ex}{\scriptsize{$t^{+}$}}\:\!
        \mathbf{D}_{#1}^{#2}\left[#3\right]}}
\newcommand{\RFI}[3]{\ensuremath{\raisebox{-.6ex}{\scriptsize{$t$}}\:\!
        \mathbf{I}_{$#1$}^{#2}[#3]}}
\newcommand{\RLD}[2]{\ensuremath{\raisebox{-.6ex}{\scriptsize{$0$}}\:\!
        \mathbf{R}_{t}^{#1}\left[#2\right]}}
\newcommand{\SIP}[3]{\ensuremath{\langle q_{#1},q_{#2}\rangle_{H^{#3}}}}
\newcommand{\Test}{\ensuremath{C^{\infty}([0,T])}}


\title{Extending Bauer's corollary to fractional derivatives}

\author{David W. Dreisigmeyer\thanks{email:dreisigmeyer@math.colostate.edu}\\
        Department of Mathematics\\
        Colorado State University, Fort Collins, CO 80523
        \and
        Peter M. Young\thanks{email:pmy@engr.colostate.edu}\\
        Department of Electrical and Computer Engineering\\
        Colorado State University, Fort Collins, CO 80523}

\maketitle
\begin{abstract}
We comment on the method of Dreisigmeyer and Young [D. W.
Dreisigmeyer and P. M. Young, J. Phys. A \textbf{36}, 8297,
(2003)] to model nonconservative systems with fractional
derivatives.  It was previously hoped that using fractional
derivatives in an action would allow us to derive a single
retarded equation of motion using a variational principle. It is
proven that, under certain reasonable
assumptions, the method of Dreisigmeyer and Young fails.\\

\end{abstract}

\section{Introduction}
\label{SEC-INTRO}

In 1931 Bauer proved the following corollary \cite{Baue31}: `The
equations of motion of a dissipative linear dynamical system with
constant coefficients are not given by a variational principle'.
There are a few methods that may allow us to get around Bauer's
corollary.  For example, we could allow additional equations of
motion to result.  Bateman used this technique in \cite{Bate31}.
If we use the Lagrangian
\begin{eqnarray}\label{BateLag}
L & = & m \dot{x} \dot{y} + \frac{C}{2} (x\dot{y} - \dot{x} y) - m
    \omega^{2} x y      ,
\end{eqnarray}
where C is a constant, we would have the following equations of
motion
\begin{eqnarray}
m \ddot{x} + C \dot{x} + m \omega^{2}x & = & 0 \label{xeom}  \\
m \ddot{y} - C \dot{y} + m \omega^{2}y & = & 0 \label{yeom}\ .
\end{eqnarray}
Bateman's method uses the loophole that Bauer's proof assumed that
no additional equations arise.

Riewe pointed out that Bauer's proof also implicitly assumes that
all of the derivatives are integer ordered \cite{Riew97}.  This
has led to attempts to use fractional derivatives in the actions
to model nonconservative systems \cite{DY03, Riew96, Riew97}. Here
we will close this second loophole by extending Bauer's corollary
to include fractional derivatives.

Our paper is organized as follows.  In section \ref{SEC-BACK} we
review the background material needed for our result.  The
extension of Bauer's corollary is proved in section
\ref{SEC-RESULT}. A brief discussion follows in section
\ref{SEC-DISC}

\section{Background material}\label{SEC-BACK}

Here we develop the relevant mathematics for our proof.  A fuller
discussion of this material can be found in \cite{DY03}.
Fractional derivatives can be defined using the theory of
distributions.  First, define the generalized functions
\begin{eqnarray}\label{PHIpdef}
\PHIp{\alpha}{t} & = & \left\{ \begin{array}{cc}
                \frac{1}{\Gamma(\alpha)}\ t^{\alpha-1} & t>0 \\
                0 & t \leq 0
                \end{array} \right.
\end{eqnarray}
and
\begin{eqnarray}\label{PHImdef}
\PHIm{\alpha}{t} & = & \left\{ \begin{array}{cc}
                \frac{1}{\Gamma(\alpha)}\ |t|^{\alpha-1} & t<0 \\
                0 & t\geq 0
                \end{array} \right.\
\end{eqnarray}
where $\Gamma(\alpha)$ is the gamma function.  The left fractional
derivatives (LFD) of a function $q(t)$ is given by
\begin{eqnarray}\label{LFDdef}
\LFD{a}{\alpha}{q} & := & \PHIp{-\alpha}{t} \ast q(t)    \\
    & = & \frac{1}{\Gamma(-\alpha)}\ \int_{a}^{t} q(\tau)
    (t-\tau)^{-(\alpha + 1)}    \nonumber
\end{eqnarray}
where we set $q(t) \equiv 0$ for $t<a$.  When $\alpha = n$, $n$ an
integer, (\ref{LFDdef}) becomes
\begin{eqnarray}\label{LFDint}
\LFD{a}{n}{q} & = & \mathcal{D}^{n} q
\end{eqnarray}
where $\mathcal{D}$ is the generalized derivative.  Right
fractional derivatives (RFDs) are defined similarly
\begin{eqnarray}\label{RFDdef}
\RFD{b}{\alpha}{q} & := & \PHIm{-\alpha}{t} \ast q(t)    \\
    & = & \frac{1}{\Gamma(-\alpha)} \int_{t}^{b} q(\tau) (\tau -
    t)^{-(\alpha + 1)}     \nonumber
\end{eqnarray}
where now $q(t) \equiv 0$ for $t>b$.  Instead of (\ref{LFDint}),
we have
\begin{eqnarray}\label{RFDint}
\RFD{b}{n}{q} & = & (-1)^{n} \mathcal{D}^{n} q\ .
\end{eqnarray}

In \cite{DY03} the actions were treated as Volterra series.  The
Volterra series are a generalization to functionals of the power
series of a function.  For a functional $\mathcal{V}[q]$, define
the symmetric kernels
\begin{eqnarray}\label{Knsymdef}
\K{n}{s}{\tau_{1},\ldots,\tau_{n}} & := & \frac{\delta^{n}
    \mathcal{V}[q]}{\delta q(\tau_{1})\cdots \delta q(\tau_{n})}\
    .
\end{eqnarray}
Now introduce the notation
\begin{eqnarray}\label{stardef}
\Kstar{n}{s}{n} & := & \int_{\tau_{1}} \cdots \int_{\tau_{n}}
    \K{n}{s}{\tau_{1},\ldots,\tau_{n}} q(\tau_{n}) \cdots
    q(\tau_{1}) d\tau_{n}\cdots d\tau_{1}\ .
\end{eqnarray}
Then $\mathcal{V}[q]$ can be written as
\begin{eqnarray}\label{Vseries}
\mathcal{V}[q] & = & \sum_{n=1}^{\infty} \frac{1}{n!}\
        \Kstar{n}{s}{n}
\end{eqnarray}
where we set $K_{0}^{(s)} \equiv 0$.

The $\Phi^{\pm}_{\alpha}(t)$ are now treated as kernels in a
Volterra series.  We can then take the functional derivative of
the series to derive our equations of motion.  An example should
make this clearer.  We will restrict our attention to the action
\begin{eqnarray}\label{ex1}
\mathcal{V}[q] & = & \frac{1}{2} K_{2} \star q^{2}
\end{eqnarray}
where $K_{2}(t,\tau)$ in (\ref{ex1}) is an arbitrary kernel, i.e.,
not necessarily symmetric as in (\ref{Knsymdef}). (Equation
(\ref{ex1}) would be sufficiently general to handle the
nonconservative harmonic oscillator.)  Now let $K_{2}(t,\tau)$ be
given by
\begin{eqnarray}\label{K2def}
K_{2}(t,\tau) & := & m\PHIp{-2}{t-\tau} + m C
    \PHIp{-\gamma}{t - \tau} + m \omega^{2} \PHIp{0}{t-\tau}
\end{eqnarray}
where $0 < \gamma < 2$ and $C$ is a constant.  So (\ref{ex1})
becomes
\begin{eqnarray}\label{ex2}
\mathcal{V}[q] & = & \frac{m}{2}\ \int_{t}\int_{\tau} \left[
    \PHIp{-2}{t-\tau} +  C \PHIp{-\gamma}{t - \tau} +
    \omega^{2} \PHIp{0}{t-\tau} \right] \times  \\
    & & q(\tau)q(t) d\tau dt \ .       \nonumber
\end{eqnarray}
The functional derivative of (\ref{ex2}) is
\begin{eqnarray}\label{ex3}
\frac{\delta \mathcal{V}[q]}{\delta q(\rho)} & = &
    \underbrace{\frac{m}{2}\ \int_{\tau} \left[
    \PHIp{-2}{\rho-\tau} +  C \PHIp{-\gamma}{\rho - \tau} +
    \omega^{2} \PHIp{0}{\rho-\tau} \right] q(\tau)
    d\tau}_{\mathrm{retarded}}     \\
    & & + \underbrace{\frac{m}{2}\ \int_{t} \left[
    \PHIp{-2}{t-\rho} +  C \PHIp{-\gamma}{t - \rho} +
    \omega^{2} \PHIp{0}{t-\rho} \right] q(t)
    dt}_{\mathrm{advanced}}\ .       \nonumber
\end{eqnarray}
If we require the advanced and retarded parts of (\ref{ex3}) to
vanish separately, we have
\begin{eqnarray}
m\left[\PHIp{-2}{\tau} + C \PHIp{-\gamma}{\tau} + \omega^{2}
    \PHIp{0}{\tau}\right] \ast q(\tau) & = & 0\ \mbox{(retarded)}
    \label{ex4}         \\
m\left[\PHIm{-2}{t} + C \PHIm{-\gamma}{t} + \omega^{2}
    \PHIm{0}{t}\right] \ast q(t) & = & 0\ \mbox{(advanced)}
    \label{ex5}
\end{eqnarray}
where in (\ref{ex5}) we used the fact that $\PHIp{\alpha}{t-\tau}
= \PHIm{\alpha}{\tau-t}$.  This is the method presented in
\cite{DY03} for deriving the equations of motion.

\section{The result}\label{SEC-RESULT}

In section \ref{SEC-BACK} we reviewed the procedure Dreisigmeyer
and Young proposed in \cite{DY03} for deriving a system's
equations of motion. From (\ref{ex4}) and (\ref{ex5}) we see that
two equations are actually derived: an advanced one and a retarded
one.  So this is, effectively, a generalization of Bateman's
method (see (\ref{BateLag}) -- (\ref{yeom})).  That is, extra
equations of motion are allowed to result from the action's
variation.

We desire to have a single, retarded equation of motion to result
from a variational principle.  From (\ref{K2def}) we see that the
derivative operators are always contained in the
$K_{2}(\tau_{1},\tau_{2})$ kernel.  Perhaps it is possible to use
some kernel other than the
$\Phi^{\pm}_{\alpha}(\tau_{1}-\tau_{2})$ to have a fractional
derivative arise from an action's variation? The following theorem
shows that this is not possible within the formalism presented in
\cite{DY03}.

\begin{theo}\label{TH-New}
There does not exist a $K(\tau_{1},\tau_{2})$, $\tau_{1},\tau_{2}
\in \mathbb{R}$, such that the variation of the quantity
\begin{eqnarray}\label{theo1}
\mathcal{V}[q] & = & \int K(\tau_{1},\tau_{2}) q(\tau_{1})
    q(\tau_{2}) d\tau_{1} d\tau_{2}
\end{eqnarray}
will result in $\LFD{a}{\alpha}{q}$ for $\alpha \neq 2n$, $n$ an
integer.

\noindent\verb"PROOF." The variation of $\mathcal{V}[q]$ is given
by
\begin{eqnarray}\label{theo2}
\frac{\delta \mathcal{V}[q]}{\delta q(\rho)} & = & \left[
    K(\rho,t) + K(t,\rho) \right] \star q(t)    .
\end{eqnarray}
We will assume that
\begin{eqnarray}\label{theo3}
\left[ K(\rho,t) + K(t,\rho) \right] \star q(t) & = &
    \PHIp{-\alpha}{\rho - t} \star q(t)
\end{eqnarray}
and arrive at a contradiction.  We require that (\ref{theo3})
holds for every $q(t)$.  Then we must have
\begin{eqnarray}\label{theo4}
\left[ K(\rho,t) + K(t,\rho) \right] & = &
    \PHIp{-\alpha}{\rho - t}    .
\end{eqnarray}
Interchanging $\rho$ and $t$ in (\ref{theo4}) gives us
\begin{eqnarray}\label{theo5}
\left[ K(\rho,t) + K(t,\rho) \right] & = &
    \PHIm{-\alpha}{\rho - t}    .
\end{eqnarray}
Hence, unless $\Phi^{\pm}_{-\alpha}(\rho - t)$ is symmetric in
$\rho$ and $t$, (\ref{theo4}) and (\ref{theo5}) cannot both hold.
That is, unless $\alpha = 2n$, $n$ an integer, there does not
exist a $K(\tau_{1},\tau_{2})$, $\tau_{1},\tau_{2} \in
\mathbb{R}$, such that (\ref{theo3}) holds.$\blacksquare$
\end{theo}

Theorem \ref{TH-New} shows that, in general, the fractional
mechanics formalisms presented in \cite{DY03, Riew96, Riew97}
cannot derive a single, retarded equation of motion.  In order to
overcome this difficulty, Riewe suggested approximating RFDs with
LFDs \cite{Riew96, Riew97}. Dreisigmeyer and Young showed in
\cite{DY03} that this is not a sound idea and, instead, allowed
for an extra, advanced equation of motion.  This latter technique
is not, itself, entirely satisfactory.

\section{Discussion}\label{SEC-DISC}

Theorem \ref{TH-New} shows that some revision of our concept of an
action may be in order if we desire a variational principle to
work for nonconservative systems.  How could we derive a single,
retarded equation of motion for systems?  Our result holds even if
$K_{2}(\tau_{1},\tau_{2})$ is allowed to be complex.  We would
also require that $q(\tau_{1}) = q(\tau_{2})$ for
$\tau_{1}=\tau_{2}$ in (\ref{theo1}). That is, we do not want to
employ Bateman's method, as was done in \cite{DY03}.

One possible method proposed by Tonti \cite{Tont71} (see also
\cite{ArJo76}) is to use the convolution product in our
Lagrangians.  This leads to actions of the form
\begin{eqnarray}\label{ConvAction}
\mathcal{V}[q] & = & \int K(t-\tau_{1}-\tau_{2}) q(\tau_{1})
    q(\tau_{2}) d\tau_{2} d\tau_{1} dt      .
\end{eqnarray}
This method does allow the derivation of a single retarded
equation of motion for, e.g., the driven nonconservative harmonic
oscillator.  Unfortunately, it does not seem possible to naturally
generalize Tonti's method to higher ordered potentials.  That is,
using quantities like
\begin{eqnarray}\label{ConvPot}
&\int K(t-\tau_{1}-\ldots - \tau_{n}) q(\tau_{1}) \cdots
    q(\tau_{n})d\tau_{n}\cdots d\tau_{1} dt&
\end{eqnarray}
in the action will not lead to the correct form for the potential
energy terms when $n>2$.  This situation should be contrasted with
that in \cite{DY03}.  There terms like
\begin{eqnarray}\label{DYPot}
&\int \PHIp{0}{\tau_{1}-\tau_{2}} \cdots
    \PHIp{0}{\tau_{n-1}-\tau_{n}} q(\tau_{1}) \cdots q(\tau_{n})
    d\tau_{n} \cdots q\tau_{1}
\end{eqnarray}
were able to treat the potential energy terms correctly.  However,
as Theorem \ref{TH-New} demonstrates, the formalism in \cite{DY03}
is unable to deal correctly with the fractional derivative terms.

An interesting feature of (\ref{ConvAction}) versus (\ref{theo1})
is the presence of $t$ in the action along with $\tau_{1}$ and
$\tau_{2}$.  Our stated goal is to derive purely retarded
equations of motion using a variational principle.  However, to
achieve this we need to find the correct kernels for our Volterra
series action. Theorem \ref{TH-New} tells us that kernels of the
form $K_{2}(\tau_{1},\tau_{2})$ are not sufficient for our
purposes. Equation (\ref{ConvAction}) suggests that we look
instead at the kernels $K(t,\tau_{1},\ldots,\tau_{n})$ for our
actions.

It is desirable to be able to use the same type of kernel for the
fractional derivatives as well as the higher ordered potential
terms.  This assumption is necessary so that the same perturbation
of $q(\tau)$ can be used in the kinetic and potential energy terms
of the action.  It allows us to reject using kernels of the form
$K(t-\tau_{1} - \cdots - \tau_{n})$, as in (\ref{ConvPot}).  Also,
it prevents us from using terms like (\ref{ConvAction}) for the
kinetic energy and terms like (\ref{DYPot}) for the potential
energy, within the Lagrangian formalism.  Finding the correct
kernels $K(t,\tau_{1},\ldots, \tau_{n})$ for our Volterra series
actions is a line of research that we are actively pursuing at
this time.

\section{Acknowledgements}\label{SEC-ACKNOW}
The authors would like to thank the NSF for grant \#9732986.  Our
deepest appreciations to the referee for bringing references
\cite{ArJo76} and \cite{Tont71} to our attention.


\bibliographystyle{plain}
\bibliography{FracDer}

\begin{thebibliography}{1}

\bibitem{ArJo76}
A.~M. Arthurs and M.~E. Jones.
\newblock On variational principles for linear initial value problems.
\newblock {\em Journal of Mathematical Analysis and Applications}, 54:840--845,
  1976.

\bibitem{Bate31}
H.~Bateman.
\newblock On dissipative systems and related variational principles.
\newblock {\em Physical Review}, 38:815--819, 1931.

\bibitem{Baue31}
P.S. Bauer.
\newblock Dissipative dynamical systems {I}.
\newblock {\em Proceedings of the National Academy of Sciences}, 17:311--314,
  1931.

\bibitem{DY03}
D.~W. Dreisigmeyer and P.~M. Young.
\newblock Nonconservative {L}agrangian mechanics: a generalized function
  approach.
\newblock {\em Journal of Physics A}, 36:8297--8310, 2003.

\bibitem{Riew96}
F.~Riewe.
\newblock Nonconservative {L}agrangian and {H}amiltonian mechanics.
\newblock {\em Physical Review E}, 53:1890--1898, 1996.

\bibitem{Riew97}
F.~Riewe.
\newblock Mechanics with fractional derivatives.
\newblock {\em Physical Review E}, 55:3581--3592, 1997.

\bibitem{Tont71}
E.~Tonti.
\newblock On the variational formulation for initial value problems.
\newblock {\em Annali de Matematica Pura ed Applicata}, 95:331--359, 1971.

\end{thebibliography}

\end{document}